# Single-shot optical precessional magnetization switching of Pt/Co/Pt ferromagnetic trilayers

Rui Xu,[1,*] Chen Xiao,[1,2,*] Xiangyu Zheng,[1,*] Renyou Xu,[1] Xiaobai Ning,[1] Tianyi Zhu,[3] Dinghao Ma,[1] Kangning Xu,[1] Fei Xu,[1] Youguang Zhang,[2] Boyu Zhang,[1,†] and Jiaqi Wei[1,†]

[1]School of Integrated Circuit Science & Engineering, Beihang University, Beijing 100191, China

[2]School of Electronic Information Engineering, Beihang University, Beijing 100191, China

[3]School of General Engineering, Beihang University, Beijing 100191, China

**ABSTRACT**. Ultra-fast magnetization switching triggered by a single femtosecond laser pulse has gained significant attention over the last decade for its potential in low-power consumption, high-speed memory applications. However, this phenomenon has been primarily observed in Gd-based ferrimagnetic materials, which are unsuitable for storage due to their weak perpendicular magnetic anisotropy (PMA). In this work, we demonstrated that applying a single laser pulse and an in-plane magnetic field can facilitate magnetic switching in a Pt/Co/Pt ferromagnetic trilayers stack within a specific laser power window. To further understand this phenomenon, we introduce a Cu layer to accelerates the re-establishment time of the anisotropy field of Pt/Co/Pt trilayers, which leads to bullseye-patterned magnetic switching. We have mapped state diagrams for these phenomena, and through micromagnetic simulations, we have determined that these switchings are influenced by thermal anisotropy torque, which can be modulated through PMA. These findings indicate that single-shot optical precessional magnetization reversal is feasible in a broader range of materials, opening avenues for the development of optical-magnetic memory devices.

The spin injection is an important research area in spintronics, which forms the physical foundation of magnetic recording [1]. Besides magnetic field, various mechanisms, including spin torque transfer[2], spin Hall effect [3], and electric field [4,5] or strain assisted switching [6,7], have been implemented to switch magnetization in spintronic devices. However, the magnetization switching process in all these approaches is governed by a precessional motion, as described by the thermodynamic Landau-Lifshitz-Gilbert (LLG) equation [8,9]. Their operation speed is fundamentally limited by the spin precession time, which all takes place in the tens of picoseconds or sub-picosecond time region [10,11]. Such a time constraint severely limits the development of spintronics.

---

[*] These authors contributed equally to this work.
[†] Corresponding author: boyu.zhang@buaa.edu.cn
weijiaqi@buaa.edu.cn

In 2007，Stanciu et al. proposed a new method of magnetization switching, which is faster than using magnetic field pulses [12]. demonstrated that the magnetization of GdFeCo could be reversed using circular femtosecond laser pulse, named all optical switching (AOS). Later, Radu et al. discovered that a single linearly polarized pulse is sufficient to switch the Gd-Fe-Co. This phenomenon was explained by a strong difference in demagnetization timescales between the two sublattices of transition metal (TM) and rare earth (RE) materials, indicating that it cannot be applied to ferromagnet [13,14]. However, for high density data storage applications, materials based on the TM-RE system, ex. Co-Gd system, face significant challenges, largely due to the insufficient perpendicular magnetic anisotropy (PMA) of RE and the TM system, which is required to stabilize competitively small domains [15]. Co/Pt ferromagnetic multilayers has been proved to achieve AOS under multi-pulse femtosecond laser excitation [16]. Although Co/Pt has stronger PMA than TM-RE system, which is suitable for high-density storage, its incapacity to achieve single pulse AOS results in lower writing efficiency [17]. Consequently, how to expand the single-pulse switching to more materials, especially ferromagnets, became a hot topic in recent years.

In 2019，Davis et al. [18] demonstrate a new heat-assisted route for sub-nanosecond magnetic recording, which is completely different from single-shot switching in GdFeCo. They show bullseye domain pattern in YIG through the combined action of an ultrashort laser pulse and an externally in-plane magnetic field. The laser pulse gently diminishes the temperature-dependent magnetocrystalline anisotropy, leading to the emergence of a transient effective torque, named thermal anisotropy torque (TAT) [19,20]. TAT then induces large-amplitude precession of the magnetization around the in-plane field. The cone angle of the precession is large enough to induce the magnetization, within one-half of a precessional period, to reverse the sign of its perpendicular component. Although complete toggle switching has not been achieved in their work, they provide a new path towards ultrafast magnetization control in ferromagnetic materials.

Here, we report the TAT assisted single-shot optical precessional magnetization reversal of Pt/Co/Pt PMA ferromagnetic trilayers. We demonstrate that the toggle switching obtained in Pt/Co/Pt trilayers is robust for laser pulses as fast as 50 fs and with respect to layer thicknesses. Furthermore, we insert Cu layer between Co/Pt and detailed studied how the insertion layer affects PMA by using first-principles calculations. We found that bullseye patterns tend to appear more in Co/Cu/Pt materials due to the acceleration of cooling rate in the film during thermal relaxation. Finally, through the micromagnetic simulation, we determined that single-shot optical magnetization reversal in Pt/Co/Pt trilayers is precessional, correlated by TAT. In this case, we can modulate the TAT by adjusting the PMA, which can extend the precessional optical magnetization switching to more ferromagnetic thin film materials. These research findings on the influence of in-plane fields and film characteristics on TAT contribute to a better understanding of the fundamental mechanism behind this switching process and pave the way for the application of magneto-optical storage.

To explore the potential of applying the TAT mechanism to ultrafast manipulation for ferromagnetism, we first prepare the stack of Ta (5 nm)/Pt (4 nm)/Co (0.8 nm)/Pt (3 nm) by magnetron sputtering, which exhibited good PMA. The TAT is introduced to Pt/Co/Pt trilayers by applying an external in-plane field. For observation of ultrafast laser-induced magnetization reversal, we built a Magneto-Optical Kerr effect (MOKE) wide-field microscopy, integrated with oblique incidence laser. Femtosecond laser pulses with a 50-fs duration at the central wavelength of 800 nm are applied to trigger magnetization dynamics. The laser incidence angle is 45°. The incident laser fluence (F) was calculated using the formula F = P / fS, where P is the laser power, f is the laser's repetition rate (set to 1 kHz), and S is the beam spot area. $S=\pi R^2$, with R being the radius of the beam spot, which is equal to FWHM /$\sqrt{\ln 2}$, where FWHM is half of the laser's spatial full-width half-maximum.

We successfully observed a toggle switching from red to blue for a linearly polarized femtosecond single laser pulse with a fluence of F = 4 mJ/cm² and an in-plane field of 3 kOe, corresponding to single-shot M+ to M- switching. With well-designed electromagnet, the out-of-plane components of the magnetic field can be eliminated and its influence on the magnetization process is avoided. As shown in Fig. 1(a), the switching can still be achieved under an opposite in-plane field. We confirmed the stability of single-shot magnetization reversal of Pt/Co/Pt trilayers by exciting thin film with 6 times of lower fluence laser pulses, as shown in Fig. 1(b). This striking behavior has been studied systematically as a function of the in-plane magnetic field, which allows us to draw a complete state diagram (in-plane magnetic field versus fluence) illustrated in Fig. 1(c) (see Supplemental Material Note 1). Surprisingly, single-shot optical switching of Pt/Co/Pt can be realized in a wide window of fluence versus in-plane magnetic field. Upon further increasing the in-plane field close to the magnetic anisotropy field (~0.8T), the potential barrier between the initial and the final states diminishes, and the switching would not occur.

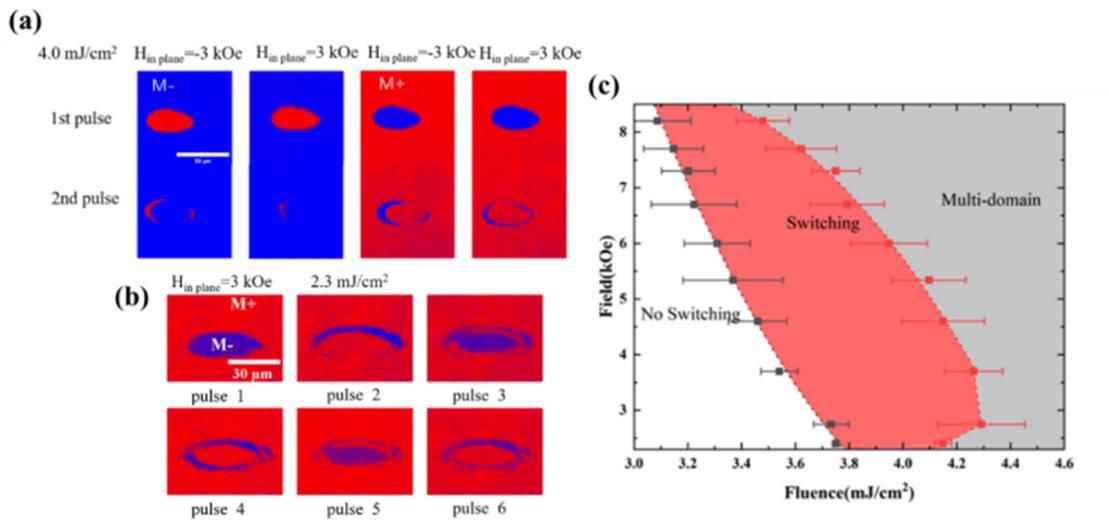

FIG. 1. Single-shot switching and state diagram for Ta(2 nm)/Pt(3 nm)/Co(0.8 nm)/Pt(3 nm) trilayers under 3 kOe in-plane magnetic field. M+ (respectively M−) corresponds to magnetization pointing perpendicular to film plane, along +z (respectively −z) in red (respectively in blue). (a) Kerr images after each single 50-fs pulse with 4 mJ/cm² laser

fluence, starting from M+ or M- state. (b) Kerr images of 6 linearly polarized laser pulses with a pulse duration of 50 fs. (c) State diagram of reversal with in-plane magnetic field versus fluence.

To investigate the influence of Co layer thickness on TAT-assisted switching, we prepared Co films with thicknesses of 0.8 nm, 1 nm, and 1.2 nm. Results show a decrease in the minimum in-plane field required for switching as the Co layer thickness is increased, as shown in Fig. 2. The results also indicate a weak dependence of Co thickness on this single-shot toggle switching.

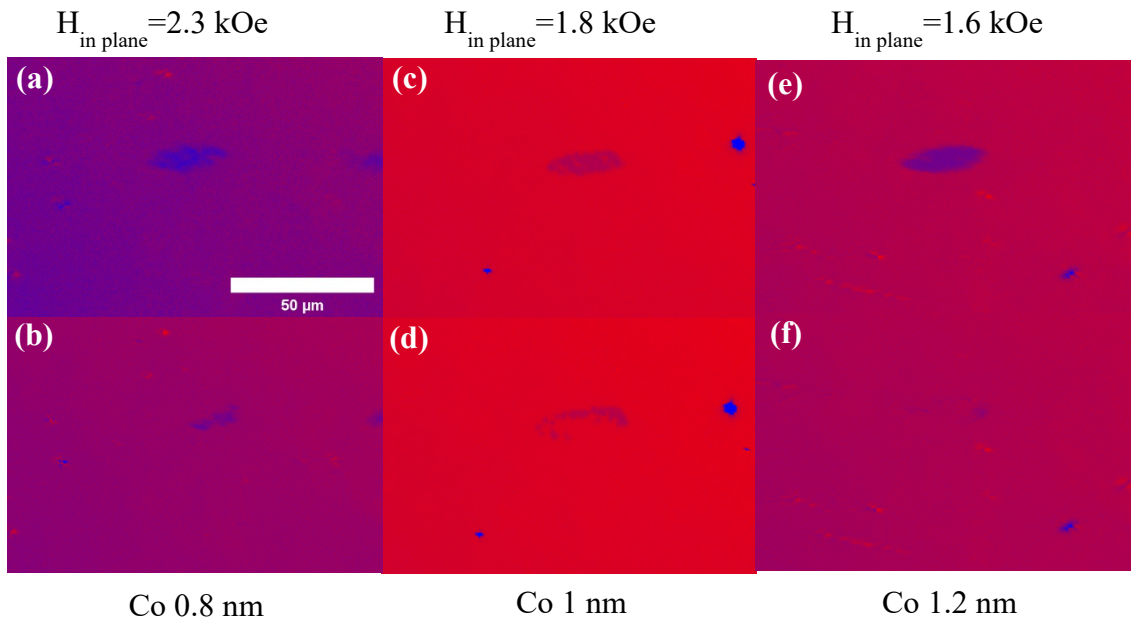

FIG. 2. Kerr images of Co with different thicknesses under the critical in-plane magnetic field. Lower in-plane magnetic field is required to reverse magnetization at thicker Co sample.

To further understand the mechanism of TAT in Pt/Co/Pt trilayers, we have systematically studied Pt/Co/Pt samples by inserting Cu layer. Films of Ta (2 nm)/Pt (3 nm)/Co (0.8 nm)/Cu (1 nm)/ Pt (3 nm) were prepared, which shows weaker PMA than previous ones, since the PMA originates only from the interfacial effects between the bottom Pt and Co. It can be seen that the TAT-assisted switching can also be achieved as shown in Fig. 3(a). Notably, a distinctive bullseye pattern emerges when the in-plane magnetic fields reach up to 5 kOe and the number of rings increases with the laser fluence, as described in Figure 3(b). Different from the results of Pt/Co/Pt trilayers film, the single domain switching can only be realized within a narrow range of laser fluence, and bullseye domain pattern appears with further increase of laser fluence. To deepen our understanding, we further preform experiments with different in-plane fields and laser powers to map the state diagram, illustrated in Figure 3(c) (see Supplemental Material Note 1). Although it is similar with that obtained from the stack with double Co-Pt interfaces, clear differences can be identified. The TAT-assisted switching takes places only within a narrow laser fluence range, and both the critical switching fluence and in-plane field are significantly reduced.

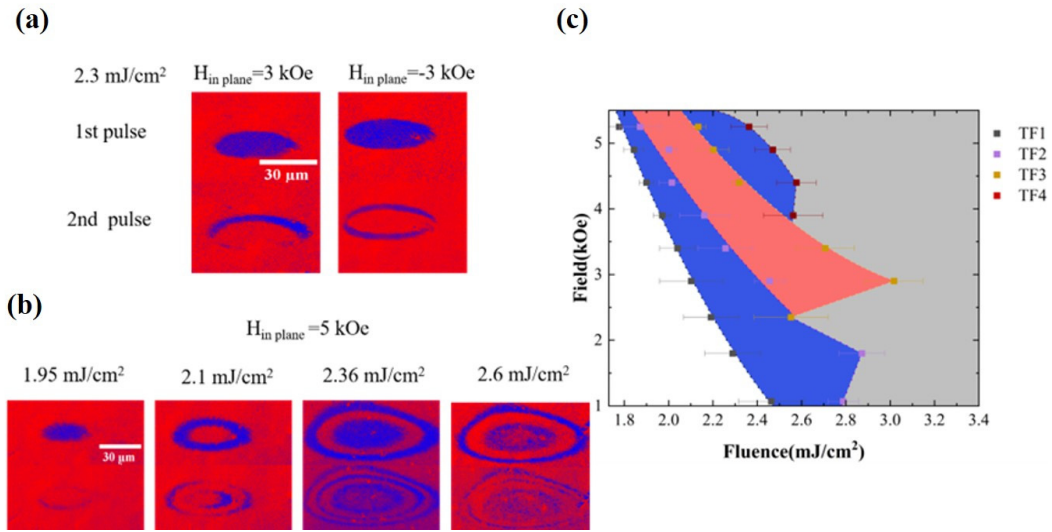

FIG. 3. Single-shot switching and state diagram for Ta (2 nm)/Pt (3 nm)/Co (0.8 nm)/Cu (1 nm)/ Pt (3 nm) trilayers under 3 kOe in-plane magnetic field. M+ (respectively M−) corresponds to magnetization pointing perpendicular to film plane, along +z (respectively −z) in red (respectively in blue). (a) Kerr images after each single 50-fs pulse with 4 mJ/cm² laser fluence, starting from M+ or M- state. (b) Kerr images of Co/Cu/Pt under different laser powers at 5 kOe in-plane field showing bullseye patterns. (c) The switching state diagram of Co/Cu/Pt system. TF1 is the boundary of critical flipping, TF2 represents when a single-layer ring begins to appear, TF3 represents when a second-layer ring begins to appear, and TF4 is the boundary of critical thermal demagnetization.

To clarify the role of field in such kind of switching, we developed a micromagnetic model based on the Landau-Lifshitz-Gilbert equation. As shown in Fig. 4, magnetic switching is realized with suitable heat stimulus under the instantaneous heating of 300K. More importantly, the final state of magnetization changes with the rising of in-plane field (see Supplemental Material Note 2).

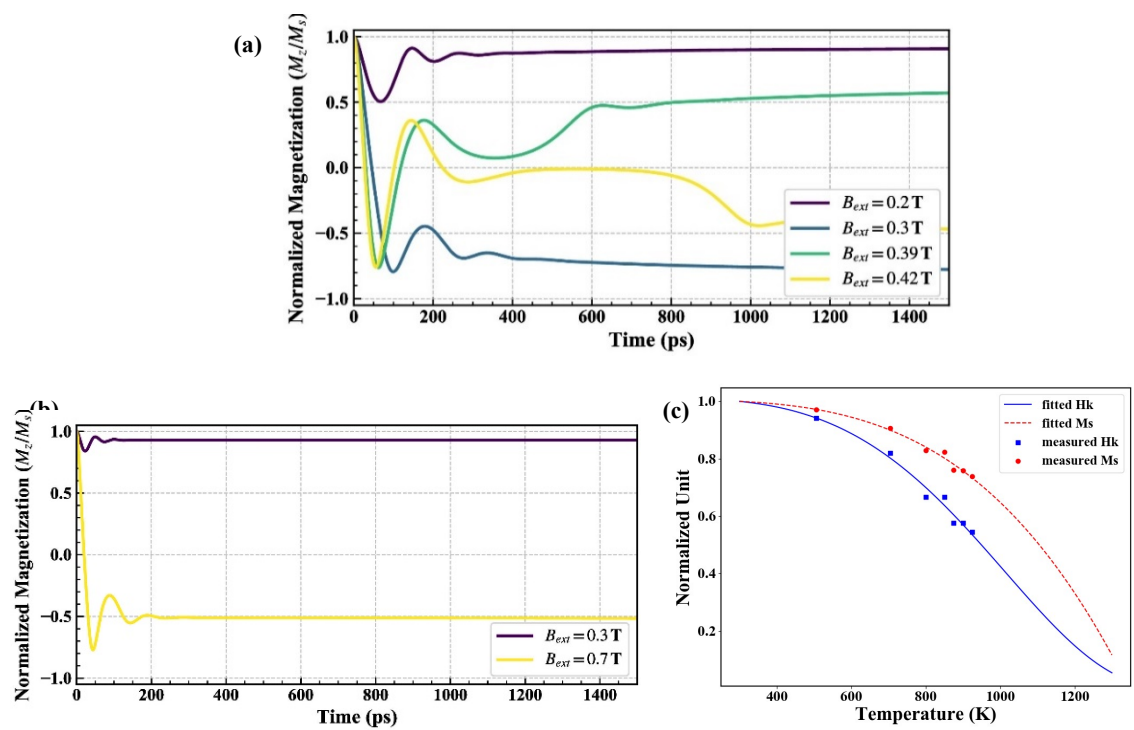

FIG. 4. (a) The simulation results for different in-plane fields of Co/Cu/Pt magnetic switching with the instantaneous heating of 300K: (b) The simulation results for different in-plane fields of Co/Pt magnetic switching with 50K. (c)The temperature-dependent results of the saturation magnetization and the magnetic anisotropy field for Co/Pt.

To further compare with experiment results, we also draw the state diagram according to the simulation results. As seen in Fig. 5b, the magnetization variation is closely related to both the temperature and the magnetic field. Moreover, the switching area exhibits periodic distribution, which is in good agreement with that shown in Fig. 3c.

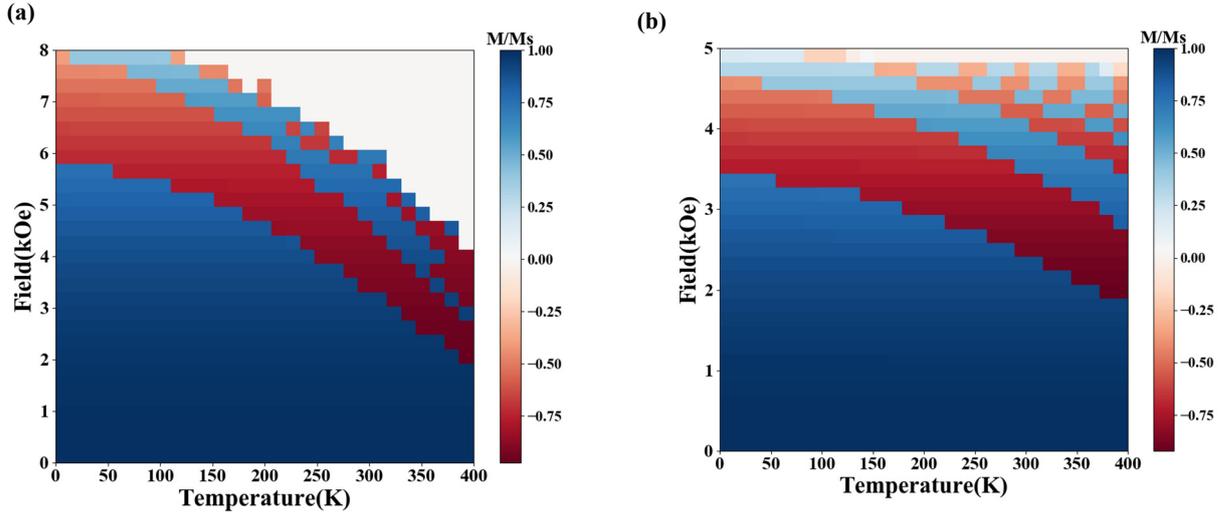

FIG. 5. (a) Co/Pt, (b) Co/Cu/Pt. The simulation results of different heating scenarios and the phase diagram of the in-plane field.

The appearance of bullseyes confirms a thermal change in the magnetic anisotropy of Co/Pt with the external in-plane field triggered by a rapid increase in the lattice temperature [18]. Referring to Ref [21], assuming that the anisotropy is modified at a sufficiently short timescale, the thermal anisotropy torque at t = 0 is expressed as:

$$\frac{1}{\gamma}T_{TAT} = \frac{1}{\gamma}\frac{dM}{dt}\bigg|_{t=0} = -M \times \Delta H_a \qquad (1)$$

In Eq. (1), as the external in-plane field strengthens along the hard axis, the equilibrium orientation of magnetization gradually changes, increasing the thermal anisotropy torque. Consequently, the observed threshold fluence for switching decreases with an increase in the in-plane field.

Additionally, we observed a higher propensity for bullseye pattern formation in the Co/Cu/Pt materials system. The incorporation of copper elevates the Curie temperature ($T_C$) of the film [22]. Crucially, copper insertion significantly accelerates the cooling rate of the film during thermal relaxation. This effect is enhanced by copper's exceptionally small absorption of near-infrared light, which minimizes heat deposition in the copper layer despite its relatively low specific heat. The resulting rapid quenching causes the magnetic anisotropy field ($H_K$) to re-establish its preferential orientation on an ultrafast timescale. Consequently, the magnetization direction becomes 'frozen' within half a precessional cycle induced by TAT. The capture of magnetic moments in these non-equilibrium intermediate states—before completing a full precession—leads to the stabilization of concentric switched/unswitched domains, thereby increasing the number of observable rings in the bullseye pattern.

Experimental data shows that the energy consumption for thermal anisotropy torque-based magnetization reversal in Co/Pt is lower than that in iron garnet magnetic thin film and comparable to that using spin-orbit torque [23]. For a 100-nm-diameter hard disk grain, the write energy is 180 fJ/bit, likely due to the thinner Co/Pt film compared to iron garnet film.

Furthermore, we have demonstrated the mechanism of TAT magnetization reversal in Co/Pt, but it could extend to other materials. We believe that similar switching via thermal anisotropy torque can occur in Co-Pt alloys and other ferromagnetic multilayers and alloys, proving that the change in magnetic anisotropy with temperature is more pronounced than that in magnetization. For instance, altering the thickness of film layers [24] or inserting additional film layers [25] can modify the temperature dependence of interlayer magnetic anisotropy.

For efficient low-power application of magnetic moments switching via TAT, avoiding the necessity for external in-plane magnetic fields during device operation is desirable. To induce magnetization precession, we can identify sources of in-plane anisotropy or in-plane exchange biasing effects, which functions similarly to an in-plane magnetic field. For instance, growth on flexible substrates induces in-plane magnetic anisotropy through strain [26], or coupling ferromagnetic and ferroelectric materials via strain can achieve similar effects [27]. IrMn or NiFe can induce an in-plane exchange bias effect on the overlying ferromagnetic material[28,29].

# Acknowledgement

The authors thank the National Key Research and Development Program of China (Grant No. 2022YFA1402604), the National Natural Science Foundation of China (Grant Nos. 12104031 and T2394473), the Fundamental Research Funds for the Central Universities (Grant Nos. YWF-23-Q-1082 and YWF-23-Q-1089) and the Support Plan for High-Level Student Science and Technology Innovation Teams from Beihang University (Grant Nos. 501XSKC2024142001 and 501XSKC2025142001) for their financial support of this work. The authors would also like to acknowledge the Center for Micro-Nano Innovation of Beihang University for the assistance in the fabrication and characterization of the device reported in this work.